\documentclass[journal]{IEEEtran}
\usepackage{cite}

\ifCLASSINFOpdf
   
   \usepackage[pdftex]{graphicx}
   \usepackage[skip=2pt]{caption}
   \usepackage{multirow}
   \usepackage{tabularew}
   \usepackage[flushleft]{threeparttable}
   
   \graphicspath{{./pdf/}{./jpeg/}{./eps/}}
  \DeclareGraphicsExtensions{.pdf,.jpeg,.png}
\else
   \usepackage[dvips]{graphicx}
   \graphicspath{{./eps/}}
   \DeclareGraphicsExtensions{.eps}
\fi
\usepackage{amsmath}
\usepackage{array}

\ifCLASSOPTIONcompsoc
 \usepackage[caption=false,font=normalsize,labelfont=sf,textfont=sf]{subfig}
\else
  \usepackage[caption=false,font=footnotesize]{subfig}
\fi

\usepackage{stfloats}
\usepackage{url}
  \usepackage{color}

\usepackage{url}
\usepackage[font=small,skip=1pt]{caption}

\begin{document}

\title{Voltage-Driven Domain-Wall Motion based Neuro-Synaptic Devices for Dynamic On-line Learning}
%
%
%

\author{\IEEEauthorblockN{Akhilesh Jaiswal*, Amogh Agrawal*, Priyadarshini Panda and Kaushik Roy\\}
\IEEEauthorblockA{School of Electrical and Computer Engineering,
Purdue University, West Lafayette, US\\
(* Equal Contributors)\\
Email: \{jaiswal, agrawa64, pandap, kaushik\}@purdue.edu}

\vspace{-2.0ex}
}
\maketitle
\pagenumbering{gobble}

\begin{abstract}

Conventional von-Neumann computing models have achieved remarkable feats for the past few decades. However, they fail to deliver the required efficiency for certain basic tasks like image and speech recognition when compared to biological systems. As such, taking cues from biological systems, novel computing paradigms are being explored for efficient hardware implementations of recognition/classification tasks. The basic building blocks of such \textit{neuromorphic} systems are \textit{neurons} and \textit{synapses}. Towards that end, we propose a leaky-integrate-fire (LIF) neuron and a programmable non-volatile synapse using domain wall motion induced by magneto-electric effect. Due to a strong elastic pinning between the ferro-magnetic domain wall (FM-DW) and the underlying ferro-electric domain wall (FE-DW), the FM-DW gets dragged by the FE-DW on application of a voltage pulse. The fact that FE materials are insulators allows for pure voltage-driven FM-DW motion, which in turn can be used to mimic the behaviors of biological spiking neurons and synapses. The voltage driven nature of the proposed devices allows energy-efficient operation. A detailed device to system level simulation framework based on micromagnetic simulations has been developed to analyze the feasibility of the proposed neuro-synaptic devices. We also demonstrate that the energy-efficient voltage-controlled behavior of the proposed devices make them suitable for dynamic on-line  in spiking neural networks (SNNs).


\end{abstract}


%
\IEEEpeerreviewmaketitle

\section{Introduction}
%
%
%
%
\IEEEPARstart{U}{nprecedented} success in miniaturization of the field-effect transistor has lead to silicon computing platforms that have become the powerhouse of today's digital revolution. As many as over 3 billion transistors \cite{bil_tran} with feature sizes approaching 7nm \cite{7nm} can be integrated on a single silicon chip. However, despite such remarkable progress, the conventional von-Neumann computing platforms fail to deliver the required energy-efficiency for certain class of problems like recognition and classification tasks. Surprisingly, the human brain shows amazing capability for recognition and classification at a fraction of power consumed by the current hardware solutions. This has lead to widespread interest in the  \textit{brain-inspired neuromorphic computing} paradigm \cite{ibm_tn}. 

Neuromorphic systems aim to mimic the computations of the human brain in order to develop novel energy-efficient computing platforms. However, there exists an inherent mismatch between the computing model for neuromorphic systems and the underlying CMOS transistor - which forms the building block of the present hardware implementations. As such, novel nano-scale devices are required that can efficiently imitate the behavior of the underlying building blocks of a neuromorphic system - the \textit{neurons} and the \textit{synapses}.

\begin{figure}[t]
\centering
\includegraphics[width=3.5in]{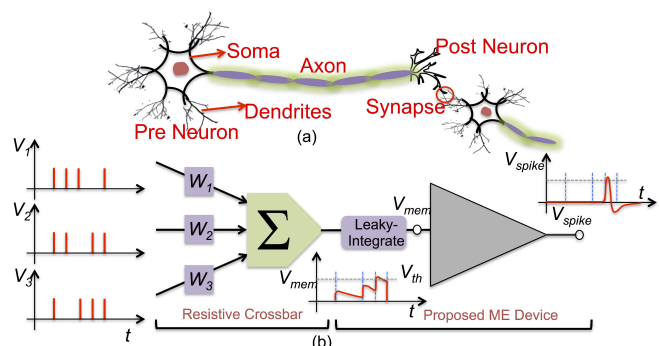}

\caption{(a) A biological neuron. Data is received by the \textit{dendrites} of the \textit{pre-neuron} in form of \textit{spikes}. When sufficient spikes are received by the pre-neuron, it emits a spike which is transmitted to the \textit{post-neuron} through the \textit{axon} and interconnecting \textit{synapses}.  }
\label{bio_neuron}
\vspace{-3mm}
\end{figure}

In Fig. \ref{bio_neuron}(a), we show a biological neuron along with the interconnecting synapses. The \textit{pre-neuron} receives electrical inputs through its \textit{dendrites} in the form of sharp electrical pulses called \textit{spikes}. Upon the reception of a particular spike the \textit{membrane-potential} ($V_{mem}$) of the associated neuron rises by a certain amount and then decays slowly until the next spike is received (see Fig. \ref{bio_neuron}(b)). When the membrane potential crosses a certain threshold ($V_{th}$), the associated neuron emits a spike. This spike is transmitted to the \textit{post-neuron} through interconnecting synapses. Thus, the neuron exhibits the leaky-integrate-fire dynamics. The spikes transmitted from one neuron to another are altered by the weights of the synaptic interconnections between the neurons. Once a neuron fires a spike, it remains non-responsive for a certain period of time known as the \textit{refractory period}. 
\begin{figure*}
\centering
\includegraphics[width=5.5in]{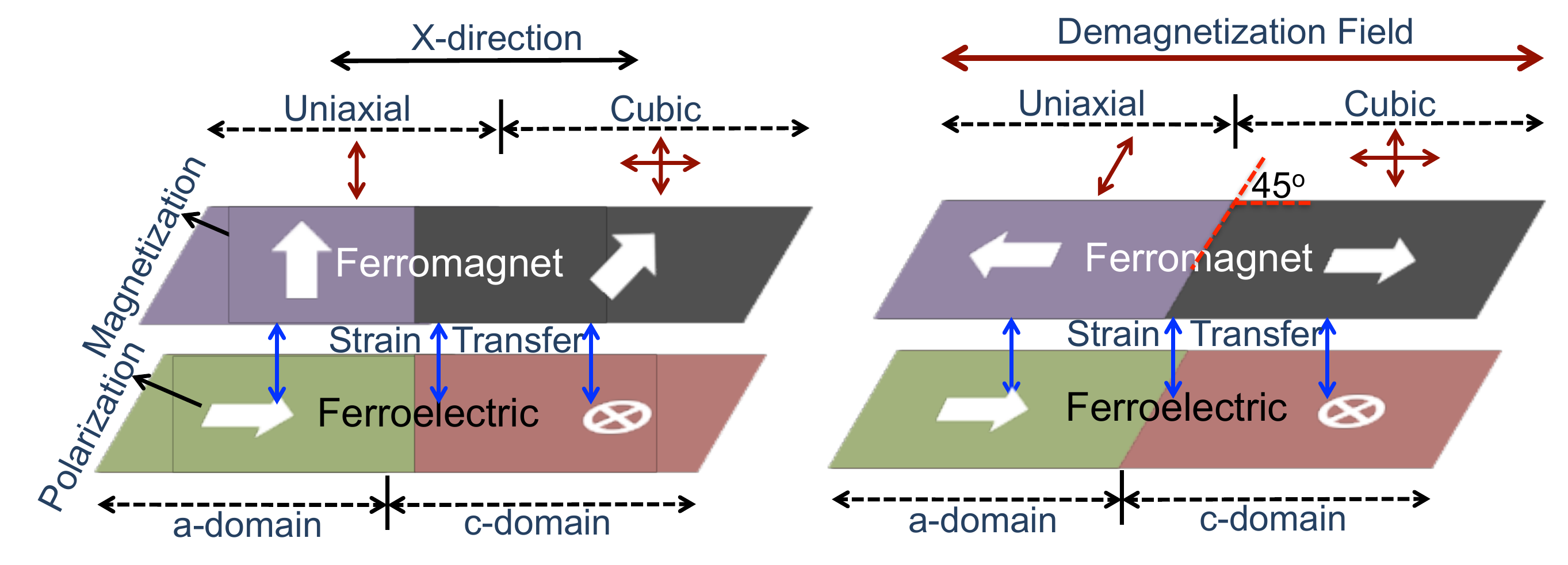}
\caption{(a) The replication of the domain pattern of the FE layer into the FM layer due to local strain coupling. An effective uniaxial anisotropy is induced in the region of the FM above the a-domain, while a cubic anisotropy is induced in the region over the c-domain. (b) Due to high aspect ratio the demagnetization anisotropy of the FM tends to align the magnetization of the FM along the length of the magnet, thereby resulting in almost 180$^o$ angle between the magnetizations in the two regions of the FM.}
\label{dw_sch}
\vspace{-3mm}
\end{figure*}
A simplified computational model of the biological neuron is shown in Fig. \ref{bio_neuron}(b). The input spikes ($V_i$) encode the information in the timing and frequency of the incoming spikes. These input spikes after being modulated by the weights ($W_i$) of the interconnecting synapses are summed up as shown in Fig. \ref{bio_neuron}(b). In response to this summation, the membrane-potential of the neuron shows a leaky-integrate behavior and emits a spike if the membrane-potential crosses the threshold, $V_{th}$. 

Hardware implementations of spiking neurons have conventionally relied on digital \cite{seo_2, modha} as well as analog \cite{ana1, ana2} CMOS circuits. Apart from area expensive implementations, CMOS spiking neurons suffer from high leakage power dissipation. Such a standby power dissipation is a major concern given that the spiking neural networks (SNNs) show large scale sparsity. Non-volatile devices that can mimic the neuronal functionality are of particular interest for such sparse systems due to zero standby power dissipation. Many non-volatile devices have been proposed as being able to exhibit the neuronal behavior. For example, phase change devices have shown to mimic the integrate-fire dynamics of biological neurons \cite{Ibm_pc}. Similarly, domain wall (DW) magnetic devices have also been reported to exhibit the integrate-fire dynamics \cite{senguptadw}. However, both the phase change and the DW neuron proposals show integrate-fire dynamics as opposed to the leaky-integrate-fire behavior of biological neurons. On the other hand, an LIF neuron based on the magnetization behavior of a ferro-magnet under an input current governed by the spin transfer torque (STT) mechanism has been presented in \cite{Abhronil}. However, the LIF characteristic of the neuron presented in \cite{Abhronil} arises due to the physical mechanisms governing the magnetization dynamics and hence, is difficult to control. Recently, a proposal for an LIF neuron using a mono-domain ferro-magnet switched by the magneto-electric (ME) effect has been reported in \cite{jaiswal2017}. Though local magnetization switching through the ME effect has been demonstrated, yet a global reversal of the magnetization vector has remained elusive \cite{ME_ramesh,heobust}. On the other hand, emerging nonvolatile memory technologies have been demonstrated for energy-efficient implementations of biological synapses including phase-change devices \cite{Suri_2011}, memristive devices \cite{Jo_2010} and spintronic devices \cite{Spin_Synapse,hybridspincmos,Srinivasan_2016}.

In this paper, we show that the recent experimental demonstrations \cite{dijken, van, van2} of a ferro-magnetic DW (FM-DW) motion through voltage driven coupling with an underlying ferro-electric DW (FE-DW) can be used to construct voltage-controlled energy-efficient LIF neuron and non-volatile programmable synapse. The key highlights of the present paper are as follows:

\begin{enumerate}
\item We propose a neuro-mimetic LIF neuron and synaptic device based on elastic coupling between an FM-DW and an FE-DW. The strong pinning of the FM-DW to the underlying FE-DW allows for pure voltage driven control of the FM-DW. The voltage driven movement of the FM-DW along with the resistance change of a magnetic tunnel junction (MTJ), allows to mimic behaviors of biological neurons and synapses. 

\item We have developed a device to circuit level simulation framework to analyze the behavior of the proposed neuro-synaptic devices. Our simulation framework comprises of micromagnetic simulation of the magnetization dynamics and non-equilibrium Green's function (NEGF) based resistance model for the MTJ.

\item The energy-efficient operation of the proposed neuro-synaptic devices stems from the voltage-controlled FM-DW motion. We demonstrate the feasibility of the proposed devices in a typical handwritten-digit recognition application for a dynamic on-line learning environment.

\end{enumerate}






\section{ Magneto-Electric DW motion based on Elastic Coupling}
\label{sec:physics}

Ferro-magnetic domain wall (FM-DW) motion has conventionally been driven by magnetic field or through the more scalable mechanism - the current induced spin transfer torque (STT) phenomenon \cite{tata}. However, the current based DW motion results in relatively high energy dissipation and therefore, extensive research investigation for pure voltage driven DW motion has gained ground in recent times \cite{dijken, van, van2}. One way of achieving such voltage driven FM-DW motion is through elastic coupling induced through a ferro-electric $-$ ferro-magnetic heterostructure \cite{van2}. Our proposal is based on the recent experimental evidence of controlled FM-DW motion under applied electric field \cite{dijken}.

The mechanism driving the FM-DW under influence of an applied electric field can be understood by referring to the heterostructure shown in Fig. \ref{dw_sch}(a). It consists of a multi-ferroic ferro-electric material like BaTiO$_3$ \cite{dijken} in physical contact with a ferro-magnet. Materials like BaTiO$_3$ show spontaneous electric polarization at room temperature. This polarization arises from the small atomic shift of Ti ions with respect to the oxygen octahedron \cite{von}. Such atomic displacement of ions with respect to one another also results in a macroscopic strain. These materials usually show stripe pattern of domains - where the displacement of constituting ions is in the same direction, separated by a thin DW. In many cases these domains are separated by an angle of 90$^o$. For example, a 90$^o$ domain wall with domains pointing in-plane (a-domains) and those pointing out-of-plane (c-domains) is shown in Fig. \ref{dw_sch}(a).

When a ferro-magnetic material is grown on top of such a ferro-electric material it experiences different amount of local strain based on the underlying domain structure of the FE material. Due to different kind of strains experienced by the FM on top of a-domain versus c-domain, different magnetic anisotropies exist in the two regions. As shown schematically in Fig. \ref{dw_sch}(a), the part of the FM over the a-domain experiences a uniaxial anisotropy while the part of the FM over the c-domain experiences a cubic anisotropy. Such different anisotropies in the region over the a- and the c-domains has been experimentally measured \cite{dijken}. The details of the crystal structure in the a- and the c- domains of the FE layer that leads to the induction of the uniaxial and the cubic anisotropy can be found in \cite{franke}.  Due to these locally induced strain anisotropy, the FM forms a domain pattern resembling the domain structure of the underlying FE material (see Fig. \ref{dw_sch}(a)). There exists a strong \textit{pinning potential} between the FE-DW and the FM-DW due to i) different amount of local anisotropies ii) the anisotropy change is almost abrupt since the FE-DW has a typical width of few lattice constants (which is an order of magnitude smaller than a typical DW width of an FM \cite{van2}). 

When a transverse electric field is applied to the FE material, the domain favoring the electric field expands at the expense of the other domain. This leads to an FE-DW motion with respect to applied electric field. Due to strong pinning, the FM-DW gets dragged with the underlying FE-DW resulting in voltage controlled FM-DW motion. One of the issues with the FM-DW formed due to elastic coupling to the FE-DW is that the FM-DW shows less than 180$^o$ difference in magnetization orientations in the two domains (refer Fig. \ref{dw_sch}(a)). This is because the magnetization of the FM over the c-domain inclines itself at an angle of 45$^o$ with the FM-DW. The position of an FM-DW can be sensed through a magnetic tunnel junction (MTJ) whose resistance varies as a function of the average magnetization direction which switches by an angle of 180$^o$. In the present case, due to voltage induced FM-DW motion, the magnetization can switch only by an angle less than 180$^o$, thereby significantly reducing the resistance sensing margin of the MTJ.

Interestingly, the FM-DW can be changed to a 180$^o$ DW by exploiting the demagnetization anisotropy of a high aspect ratio FM \cite{dijken}. Let us assume the FM on the top of the FE material is patterned into rectangular magnets such that the domain walls in the FE and FM layers make an angle of 45$^o$ with the length of the magnet, as shown in Fig. \ref{dw_sch}(a), thereby aligning the magnetization of the FM over the c-domain along the length of the FM. Due to the rectangular shape of the FM, a demagnetization anisotropy would tend to keep the magnetization of the FM aligned with the longer dimension of the FM \textit{i.e.} towards the (positive or negative) x-axis. Thereby, for the part of FM on the a-domain, the negative x-axis direction would be favored and the magnetization in that region would point in negative x-direction. Thus, through proper engineering of the FM shape one can obtain almost 180$^o$ FM-DW although the underlying FE domains exhibits a 90$^o$ separation. As we will observe later, this 180$^o$ FM-DW will allow better resistance sensing margin for the proposed devices.

\section{Magneto-Electric DW motion based Neuro-Synaptic Devices}

Exploiting the aforementioned 180$^o$ voltage driven FM-DW motion, we can construct neuro-synaptic devices, wherein the membrane-potential of the neuron and the weights of the synapses are represented by the position of the FM-DW.

\subsection{LIF Neuron}

\begin{figure}[t]
\centering
\includegraphics[width=3.2in]{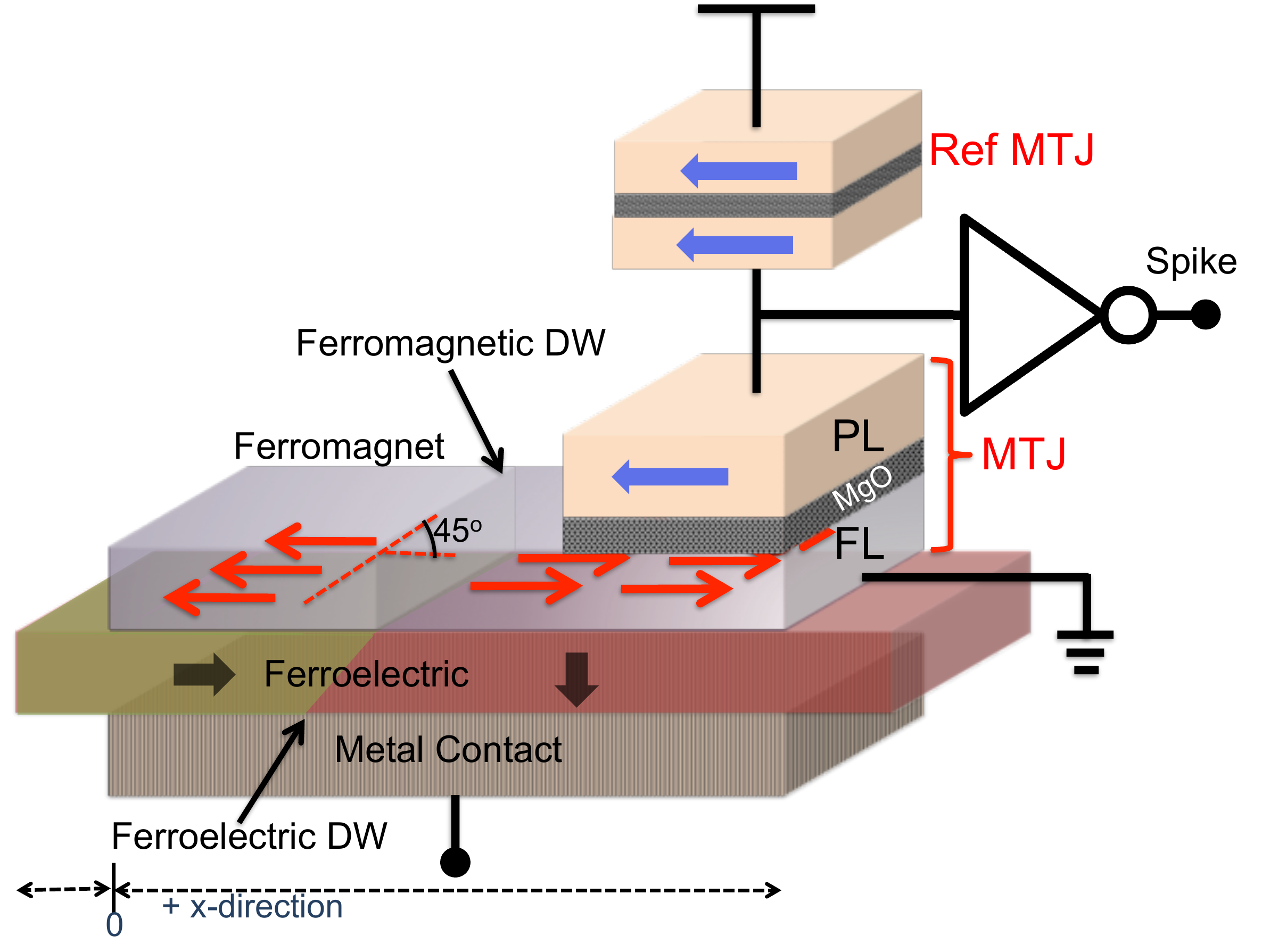}

\caption{The proposed non-volatile LIF neuron based on elastic coupling between the FE-DW and FM-DW. The position of the FM-DW represents the membrane-potential, while the switching activity of the MTJ emulates the firing behavior of the neuron.}
\label{3d}
\vspace{-3mm}
\end{figure}

The proposed LIF neuron is shown in Fig. \ref{3d}. The device consists of a ferro-magnet / ferro-electric heterostructure in contact with one another. The metal contact to the ferro-electric layer is explicitly shown in the figure. When a positive voltage is applied on the metal contact, the a-domain expands at the cost of the c-domain, resulting in an FE and FM-DW motion in the positive x-direction. Similarly, on application of a negative voltage, the c-domain expands causing a DW motion in the negative x-direction. The motion of the FE-DW drags with it the FM-DW in response to the applied voltage. The range of motion of the FE-DW is constrained by the area covered under the metal contact, thus ensuring the FE-DW never disappears in the ferro-electric material.

At the rightmost end, the ferro-magnetic layer forms an MTJ structure with a tunneling oxide (MgO) and a fixed ferromagnetic layer called the \textit{pinned layer} (PL). When the PL and the average magnetization direction (of the region of ferro-magnet under the MTJ cross-section) points in the same (opposite) direction the MTJ exhibits low resistance parallel state `P' (high resistance anti-parallel state `AP'). A reference MTJ is used to form a voltage divider connected to a CMOS inverter. The resistance of the reference MTJ and the trip point of the inverter is chosen such that the output terminal denoted as \textit{spike} goes high if and only if the lower MTJ is in parallel state. Note, the fact that the ferro-magnet shows 180$^o$ DW allows for maximum difference in the parallel and anti-parallel resistance of the MTJ, thereby increasing the voltage difference at the input of the inverter allowing robust operation.

The desirable characteristics of the proposed device can be enumerated as 1) the voltage driven motion of the DW allows low energy consumption 2) the write path (through the ferro-electric metal contact) and the read path (through the voltage divider) ensures decoupled read/write operations, thereby allowing independent optimization of the read and write paths 3) the voltage divider circuit along with the CMOS inverter allows for low overhead reading of the spiking event.

The LIF characteristic of the proposed device can be understood as follows: 1) A positive voltage on the metal contact will result in motion of the FM-DW in +x-direction. Since the DW position is non-volatile the forward motion of the FM-DW mimics the \textit{integrate dynamics} of biological neurons. 2) A small negative voltage on the metal contact would lead to an FM-DW motion in the negative x-direction, thereby imitating the \textit{leaky behavior} of the neuron. Note, it is required that the membrane-potential of a neuronal device (represented by the position of FM-DW) should keep leaking at all times except when it has received sufficient excitation in form of input spikes. For a typical current driven FM-DW motion, such a leaky characteristics would incur unacceptable energy consumption. This is due to the fact that all the neurons will continuously require a negative current flow through the device to ensure the FM-DW keeps moving slowly in the negative x-direction. In the proposed voltage controlled neuronal device, the leaky behavior merely requires a small negative voltage on the metal contact. Since ferro-electric materials are usually insulators, this negative voltage would not incur any short circuit leakage current thus allowing negligible energy overhead for mimicking the leaky behavior. 3) Finally, when the FM-DW travels far enough in the +x-direction the average magnetization of the FM under the MTJ switches and the output of the inverter goes high indicating the \textit{firing event} of the proposed device. Thus, the proposed device exhibits the non-volatile leaky-integrate-fire dynamics.

\begin{figure}[t]
\centering
\includegraphics[width=0.4\textwidth]{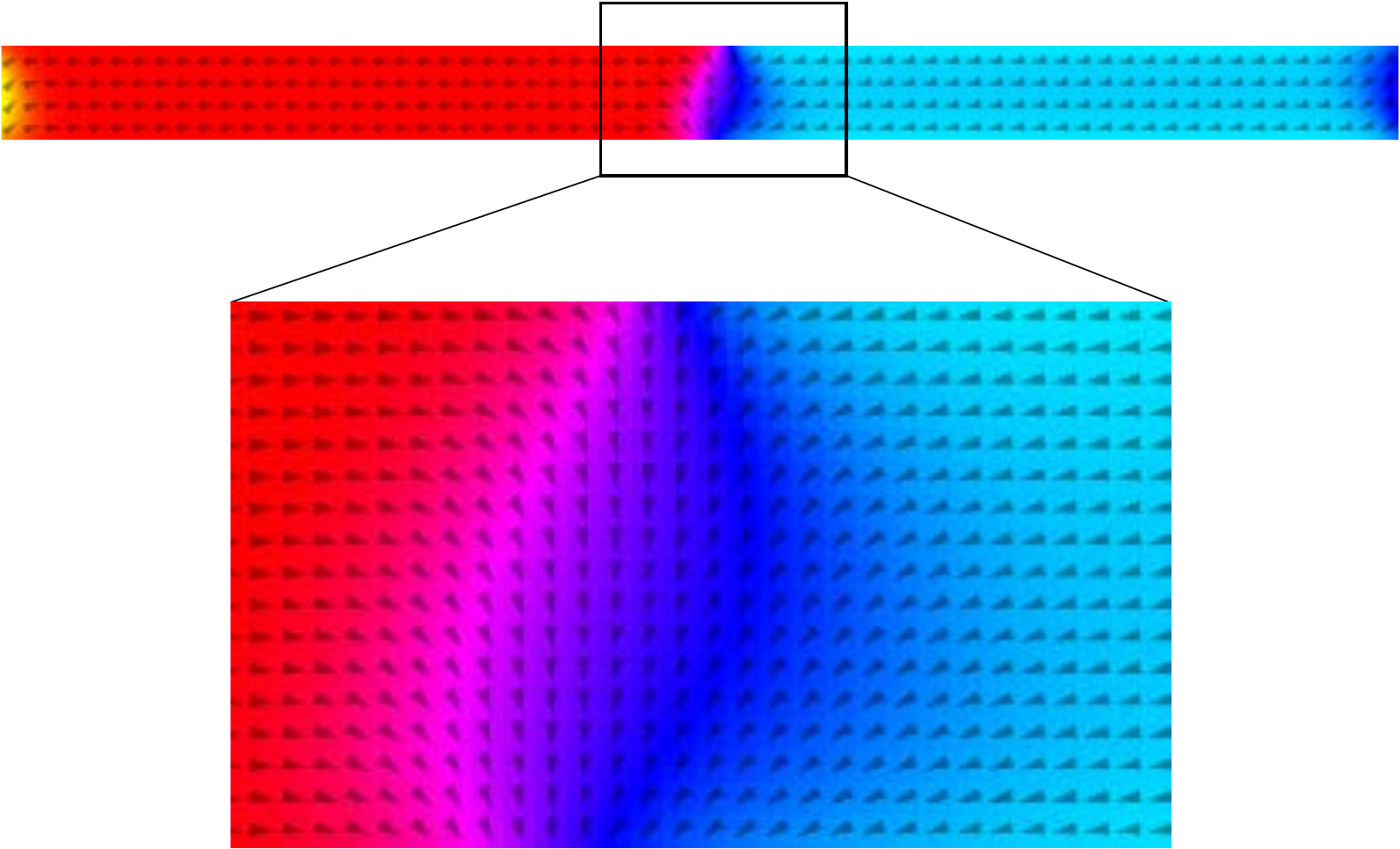}

\caption{Micromagnetic simulation showing the domain wall shape and structure. The zoomed image shows a 90$^o$ domain wall which has been transformed to a 180$^o$ domain wall due to shape anisotropy.}

\label{fig:DWzoom}
\vspace{-3mm}
\end{figure}

\subsection{Programmable Synapse}

\begin{figure}[t]
\centering
\includegraphics[width=3.2in]{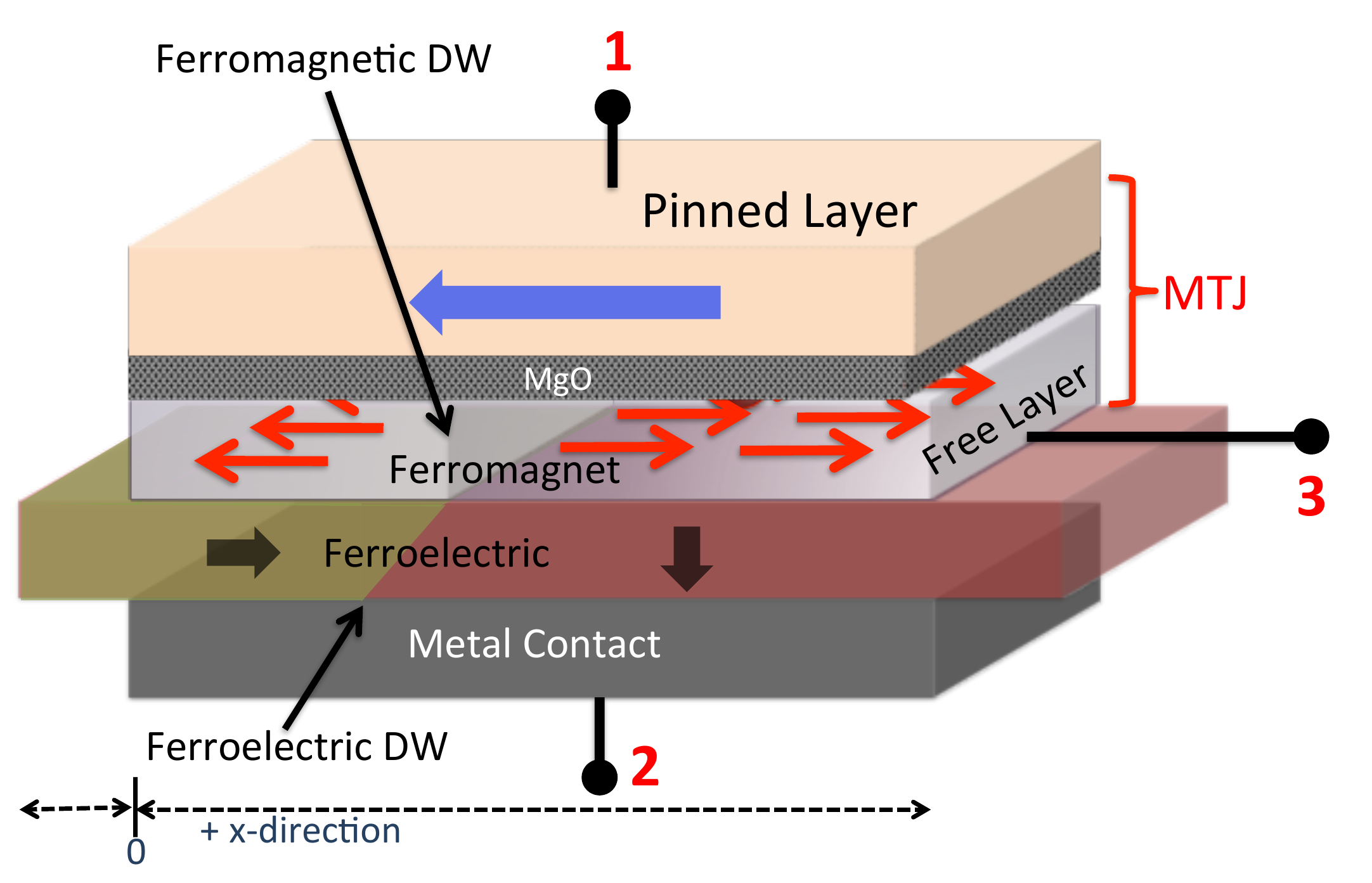}
\caption{The proposed non-volatile programmable synapse based on elastic coupling between the FE-DW and FM-DW. The position of the FM-DW modulates the conductane between Terminal-1 and 3 of the device. The FM-DW position, and thus the conductance of the synapse, can be modified by applying a +ve or -ve voltage across Terminal-2 and 3.}
\label{syn}
\vspace{-3mm}
\end{figure}

The proposed synaptic device under investigation is shown in Fig. \ref{syn}. This device is very similar to the neuronal device described above. The device consists of a ferro-magnet / ferro-electric heterostructure elastically coupled together. When a positive voltage is applied on the metal contact (between Terminal-2 and 3), the FE-DW moves in the positive x-direction. The FE-DW drags along with it the FM-DW in the positive x-direction in response to the positive voltage. Similarly, for a negative voltage, the FE-DW, and thus FM-DW, move in the negative x-direction. In addition, the free ferro-magnetic layer forms an MTJ structure with a tunneling oxide (MgO) and a pinned ferromagnetic layer (between Terminal-1 and 3). When the FM-DW position is at one end ($x=0$), the MTJ is fully in the high resistance AP state. When the FM-DW is at the other end of the device, the MTJ is in the low resistance P state. However, if the FM-DW is somewhere in between, the resistance of the MTJ is a parallel combination of AP and P, as follows:
\begin{equation}
G_{MTJ} = \frac{x\times G_{P} + (L-x)\times G_{AP}}{L}
\label{eqn:mtjcond}
\end{equation}
where $G_{MTJ}$ is the MTJ conductance, $G_P$ and $G_{AP}$ are the parallel and anti-parallel conductances of the MTJ, respectively, $x$ is the FM-DW position and $L$ is the total length of the magnet. This simplified equation holds because the length of the domain wall is small compared to the length of the magnet. The resistance or conductance can be sensed between Terminal-1 and 3. Thus, the conductance of this device can be set anywhere between $G_P$ and $G_{AP}$, representing the synaptic conductance in SNNs.

The synaptic behavior of the device can be understood as follows: 1) A positive voltage on the metal contact will result in FM-DW motion in +ve x-direction, and thereby increase the MTJ conductance. This mimics the \textit{long-term potentiation}, or strengthening of the synaptic weights in SNNs. 2) A negative voltage on the metal contact results in FM-DW motion in -ve x-direction, and decreases the synaptic conductance. This mimics the \textit{long-term depression}, or reduction in synaptic strength in SNNs. 3) Interestingly, the proposed device can exhibit leaky-behavior in addition to the usual non-volatile multi-level memory characteristics that can be used to model \textit{`forgetting'} in synapses. Recently, the authors in \cite{aspanda,arxivasp} proposed a novel learning mechanism termed as, Adaptive Synaptic Plasticity (ASP), that augments traditional neural systems with a key ability of \textit{`learning to forget'} for robust and continuous learning in a dynamically evolving environment. We use the proposed device's leaky behavior to implement the ASP learning algorithm and demonstrate the effectiveness of our proposed devices for real-time on-device learning without catastrophic forgetting. For a typical current driven FM-DW motion, such a leaky characteristic would incur unacceptable energy consumption. In the envisioned voltage controlled synaptic device, the leaky behavior merely requires a small negative voltage across the FE layer. The small negative voltage on the metal contact recedes the FM-DW position, thereby reducing the synaptic strength continuously over time.


\section{Device Modeling and Simulation}
\label{sec:simframe}
A mixed-mode simulation framework was developed for the analysis of the proposed device structure. The simulation framework consists of three components - a) the exponential dependence of FE-DW velocity on the applied voltage in accordance to the Merz's Law \cite{Merz's_law}, b) the micromagnetic response of the FM-DW due to elastic coupling with the underlying FE-DW motion, c) the resistance change of the MTJ as a function of device dimensions and magnetization directions based on non-equilibrium Green's function (NEGF) formalism \cite{knack}.

\paragraph{FE-DW velocity} The field driven dynamics of FE domain walls has been extensively studied in the past \cite{Stadler1963NucleationKVcm}, \cite{Stadler1966Forward3}, and the velocity of FE-DW has been observed to depend exponentially on the applied voltage.  This exponential dependence is given by the Merz's Law \cite{Merz's_law}. Experimental evidence of exponential FE-DW motion in BaTiO$_3$ has been demonstrated in \cite{ShinLETTERSMotion}. Merz's Law can be written as
\begin{equation}
v_{FE} = K_{FE} \times exp({a/E})
\label{merz}
\end{equation}
where $v_{FE}$ is the FE-DW velocity, $E$ is the electric field, given by $\frac{V_{FE}}{t_{FE}}$, where $V_{FE}$ is the applied voltage across the FE layer, and $t_{FE}$ is the thickness of the FE layer. $K_{FE}, a$ are fitting parameters from experimental data adopted from \cite{ShinLETTERSMotion}. As shown later in the manuscript, the FM-DW would closely follow the motion of the FE-DW \cite{van2}. As such, the FM-DW will also have an exponential dependence of velocity with respect to the applied voltage.

\begin{table}[t]
\renewcommand{\arraystretch}{1.3}
\centering
\caption{Parameters used for simulations adopted from \cite{dijken, van2}}
\label{tab:parameters}

\begin{tabular}{c c}
\hline \hline
\bfseries Parameters & \bfseries Value\\
\hline
Magnet Length ($L_{mag}$) & $ 1.5 um  $\\
Magnet Width ($W_{mag}$) & $ 100 nm $\\
Magnet Thickness ($t_{mag}$) & $ 2.5 nm $\\
FE Oxide Thickness ($t_{FE}$) & $ 100 nm$\\
Saturation Magnetization ($M_{S}$) & $1.7\times10^6 A/m$ \\
Gilbert Damping Factor ($\alpha$) & 0.01 \\
Exchange Stiffness ($K_{ex}$) & $2.1\times10^{-11} J/m$ \\
Cubic Anisotropy ($K_{c}$) & $4\times10^{4} J/m^{3}$ \\
Uniaxial Anisotropy ($K_{u}$) & $2\times10^4 J/m^{3}$ \\
Simulation cell size ($dx,dy,dz$) & $2.93,2.93,2.5 nm$ \\
Temperature ($T$) & $ 300 K$ \\

\hline \hline

\end{tabular}
\vspace{-5mm}
\end{table}

\paragraph{Micromagnetic FM-DW dynamics}  
For simulating the FM-DW dynamics in response to the underlying FE-DW motion, we used a GPU based micromagnetic simulator called MuMax \cite{mumax}. 
The a- and c-domains of the FE material result in different amount of strains due to the multi-ferroic nature of materials like BaTiO$_3$ \cite{franke}. This strain is transferred to the FM layer on top of the FE layer. Thus, the FM layer experiences different amount of strains depending on the fact whether the FE layer underneath has an a-domain or a c-domain. This results in local strain anisotropy experienced by the FM layer. When the FE-DW moves (in accordance to the Merz's Law), the local strain anisotropy experienced by the FM layer moves as well. Thus, in order to mimic the different anisotropies experienced by the FM layer, we divided the simulation region in two, each having a different strain anisotropy due to the underlying a- and c-domains. This simulation methodology is similar to the ones adopted in \cite{dijken, van2}. A uniaxial anisotropy with a constant  $K_u$ was added for that part of the FM layer that was supposed to be above the a-domain. Similarly, a cubic anisotropy with a constant $K_c$ was added for the region of the FM layer corresponding to the c-domain. Thus, there exists an anisotropy boundary (AB) in the FM layer wherein the anisotropy changes from uniaxial to cubic. The various simulation parameters used in our framework are summarized in Table I.  Recall, because of the demagnetization field one would expect the FM-DW to exhibit $\sim$ 180$^o$ DW, even though the underlying FE layer shows a 90$^o$ DW. This can be confirmed by the micromagnetic simulation results shown in Fig. \ref{fig:DWzoom}. As seen in the figure, the magnetizations near the DW have 90$^o$ difference in their orientations. But, as one moves away from the DW, the magnetizations (in the `blue' region) slowly tend to orient themselves to 180$^o$ due to the effect of the demagnetization field.

\begin{figure}[t]
\centering
\includegraphics[width=0.4\textwidth]{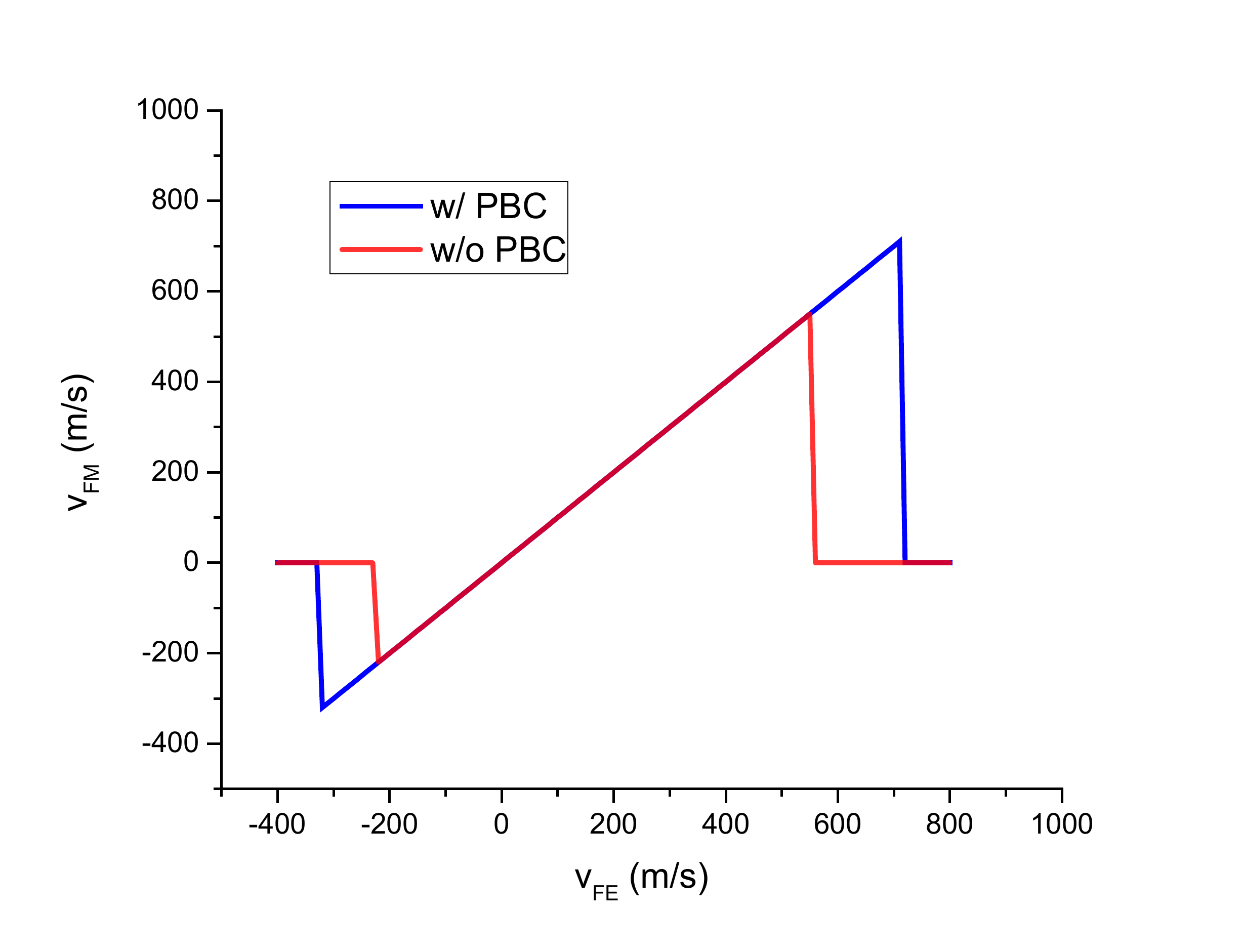}
\caption{Depinning velocities of the magnetic domain wall, for positive and negative velocities. The blue plot was obtained by using periodic boundary conditions and parameters from \cite{van2}. The red plot was obtained without periodic boundary conditions and scaled dimensions.}
\centering
\label{fig:depinning}
\vspace{-3mm}
\end{figure}

In order to mimic the movement of the FE-DW on application of a voltage across the FE layer, we shifted the AB in accordance to the Merz's law. It was found that the FM-DW followed the FE-DW linearly up to a certain velocity called the \textit{depinning velocity}. Beyond the depinning velocity, the FM-DW was not able to keep pace with the fast moving FE-DW. Thus, the maximum achievable velocity constraint arises due to the slower response of the FM-DW to the fast moving FE-DW. Note, a detailed description of the FM-DW dynamics beyond the depinning velocity can be found in \cite{van2}. In this work, we would only utilize the velocity regime below the depinning velocity where the FM-DW linearly follows the FE-DW. As shown in Fig. \ref{fig:depinning}, we plot the FM-DW velocity ($v_{FM}$) as a function of the FE-DW velocity $v_{FE}$. The blue line, represents the FM-DW velocity for a periodic boundary condition similar to \cite{van2}. On the other hand, the red line corresponds to a more realistic simulation where the magnet dimension was taken to be 1.5$um$ $\times$ 100$nm$ $\times$ 2.5 $nm$ . The simulations show a slight decrease in the depinning velocities ($\sim550$ m/s and $\sim210$ m/s respectively, for positive and negative direction), as compared to the periodic boundary case. This can be attributed to the increased demagnetization due to shape anisotropy in the smaller magnets. A higher demagnetization field suppresses the control of the strain anisotropy from the underlying FE layer. For the LIF neuron, we shall use the positive velocity for the integrate operation and the negative velocity for imitating the leaky dynamics of the neuron. Similarly, for the synaptic device, we would use the positive velocity for long-term potentiation and the negative velocities for long-term depression and leaky-behavior of synaptic weights.

\begin{figure}[t]
\centering
\includegraphics[width=0.3\textwidth]{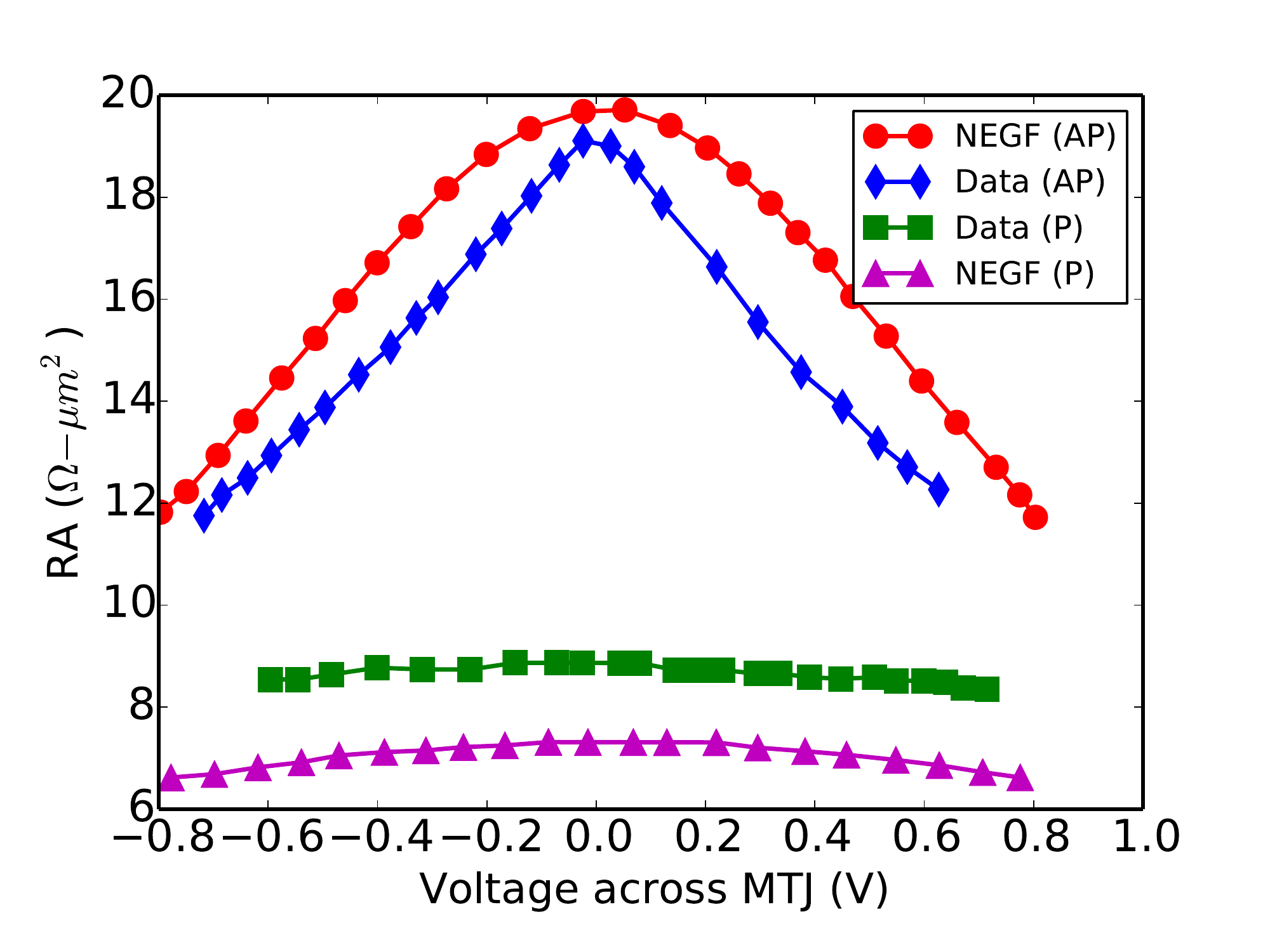}

\caption{The NEGF based resistance model \cite{knack} benchmarked against experimental data from \cite{lin2}.}

\label{fig:negf}
\vspace{-3mm}
\end{figure}

\paragraph{Resistance Model} To model the resistance of the MTJ stack, non-equilibrium Green's function (NEGF) formalism \cite{knack} was used. The details of the NEGF model for estimation of the resistance of the MTJ as a function of applied voltage and the average magnetization direction was adopted from \cite{knack}. The model was benchmarked against experimental data, as shown in Fig. \ref{fig:negf}. The results from the NEGF equations were abstracted into a Verilog-A model, which was used in SPICE simulations along with predictive technology models (PTM) \cite{PTM} for CMOS transistors.

\begin{figure}[!h]
\includegraphics[width=0.5\textwidth]{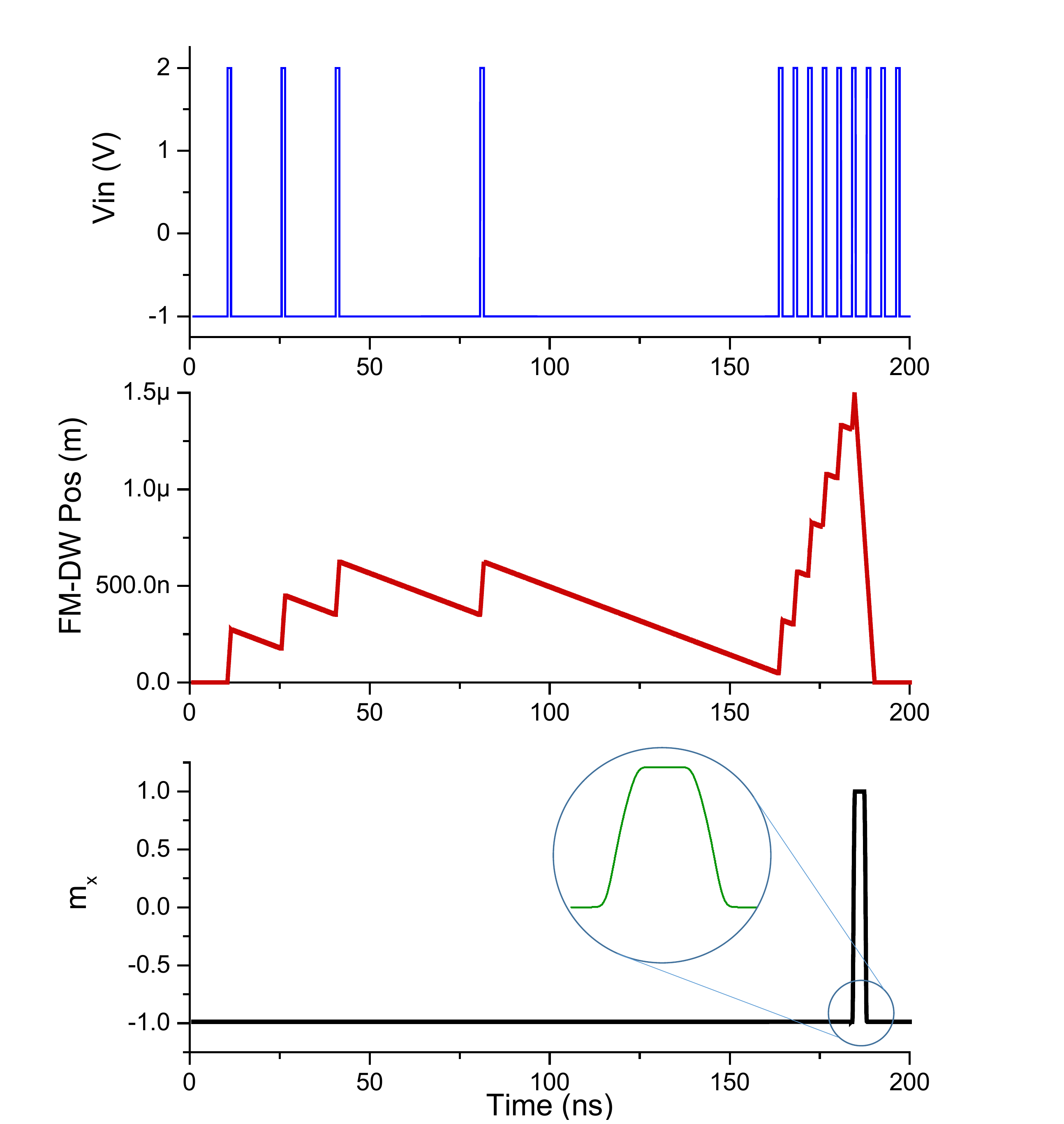}
\caption{Leaky integrate and fire behavior of the proposed neuron in response to input train of spikes. (a) Input voltage spike train received by the neuron. (b) FM-DW position (acts as membrane potential variable). (c) x-component of magnetization under the MTJ stack. Once the MTJ switches, the neuron fires, and the domain wall is reset to its initial position. The inset shows the average magnetization under the MTJ when the domain wall traverses the MTJ.}
\centering
\label{fig:firing}
\vspace{-3mm}
\end{figure}

\section{Results}

\subsection{Neuro-synaptic behavior of the proposed devices}

Fig. \ref{fig:firing} shows the behavior of the proposed neuron, obtained from the mixed-mode simulation model developed for the device. The FM-DW was initially assumed to be at $x=0$ position (refer Fig. \ref{3d}). Correspondingly, at the extreme right end of the magnet, the MTJ is in anti-parallel resistance state. 
A train of voltage spikes is applied to the metal contact, which mimics the pre-synaptic spikes received by the neuron (Fig. \ref{fig:firing}(a)). In a typical neuromorphic system, the input data is encoded in the frequency or the timing of the incoming spikes. It is to be observed that the resting potential of the input spikes is a negative voltage (-1V) upon which the spikes are superimposed and are represented by voltage pulses of amplitude 2V with a time duration of 1ns. 

When an incoming spike is applied to the neuron, the FE-DW and the FM-DW move in the positive x-direction thus implementing the integrate behavior. In absence of any incoming spike, the neuron sees a negative voltage on its input terminal and the FE-DW as well as the FM-DW slowly move towards the negative x-direction, thereby mimicking the leaky behavior (Fig. \ref{fig:firing}(b)). When the FM-DW reaches far enough in the positive x-direction, the average x-component of the magnetization ($m_x$) beneath the MTJ switches from -1 to +1 (Fig. \ref{fig:firing}(c)). In response, the output of the voltage divider would switch from low to high indicating that the neuron has emitted a spike. This output spike voltage can be used to trigger the reset phase, and a negative voltage can then be applied at the metal contact, thereby moving the domain wall back to the initial position ($x=0$). The device remains non-responsive to any more incoming spikes during this duration, mimicking the \textit{refractory period} of the neuron.

\begin{figure}[t]
\centering
\includegraphics[width=0.5\textwidth]{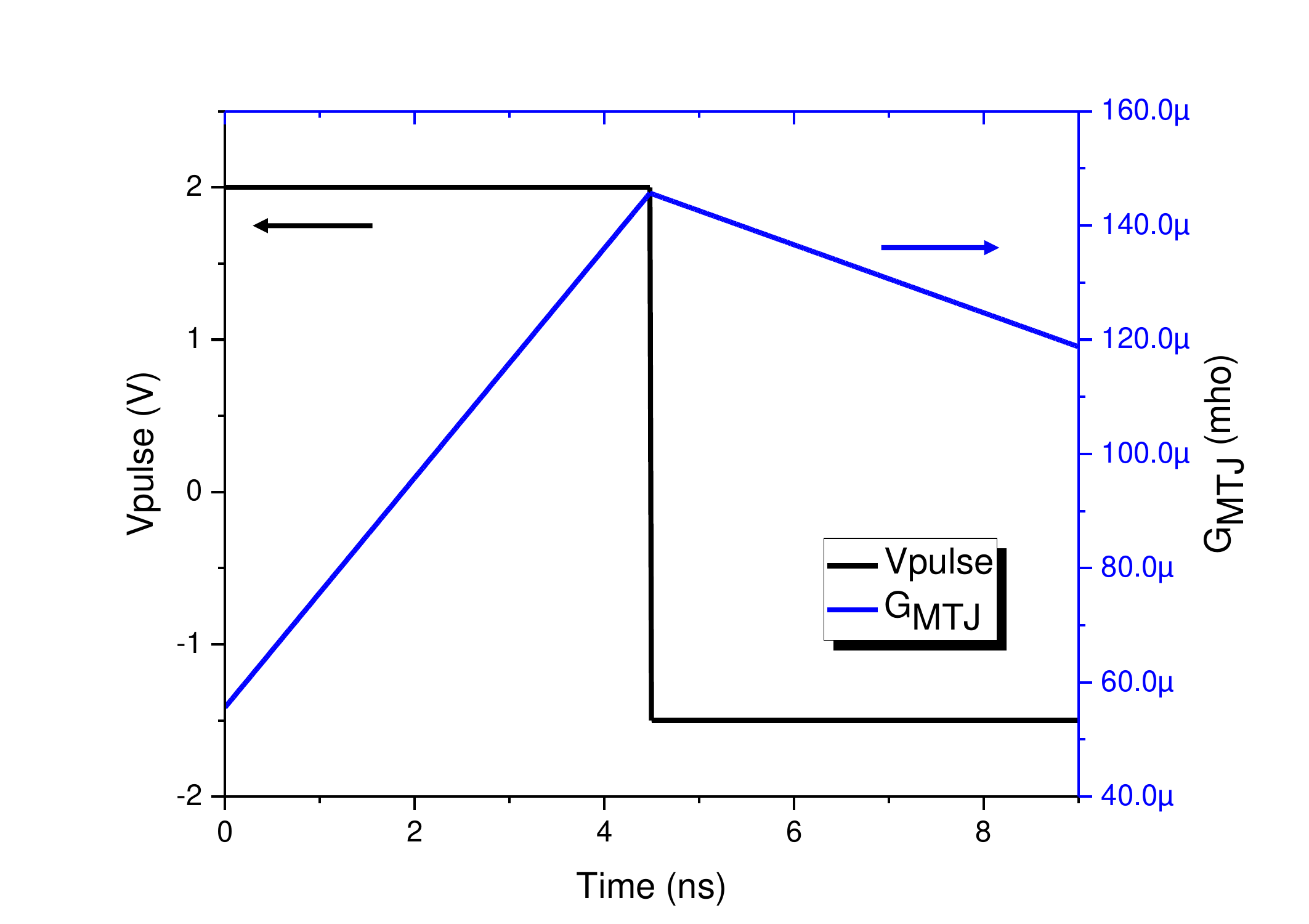}
\caption{Plot of MTJ conductance $G_{MTJ}$ of the synaptic device in response to voltage pulses exhibits a controlled behavior of the synaptic weights. This can be used for better learning algorithms like `ASP' for precise tuning of synaptic weight values. The leaky behavior of the synaptic weights can be implemented using a small -ve voltage across the device.}
\label{syn_plot}
\vspace{-3mm}
\end{figure}

In Fig. \ref{syn_plot}, we show the simulation results for the synaptic device. Again, the FM-DW was initially assumed to be at $x=0$ position (refer Fig. \ref{syn}). The conductance of the synaptic device ($G_{MTJ}$) is plotted with time in response to a positive and negative voltage pulses. During the positive applied voltage, the FM-DW moves in the +ve x-direction, thereby increasing the conductance, as described in Eqn \ref{eqn:mtjcond}. On application of a negative voltage, the FM-DW moves in the -ve x-direction, thereby decreasing the conductance. Thus we obtain a voltage-controlled conductance to set the synaptic weights as desired. Also, a small negative voltage would allow slow decay of the synaptic conductance to account for the `forgetting' mechanism in synapses.

\subsection{Dynamic Learning in SNNs}

\begin{figure}[t]
\centering
\includegraphics[width=3.2in]{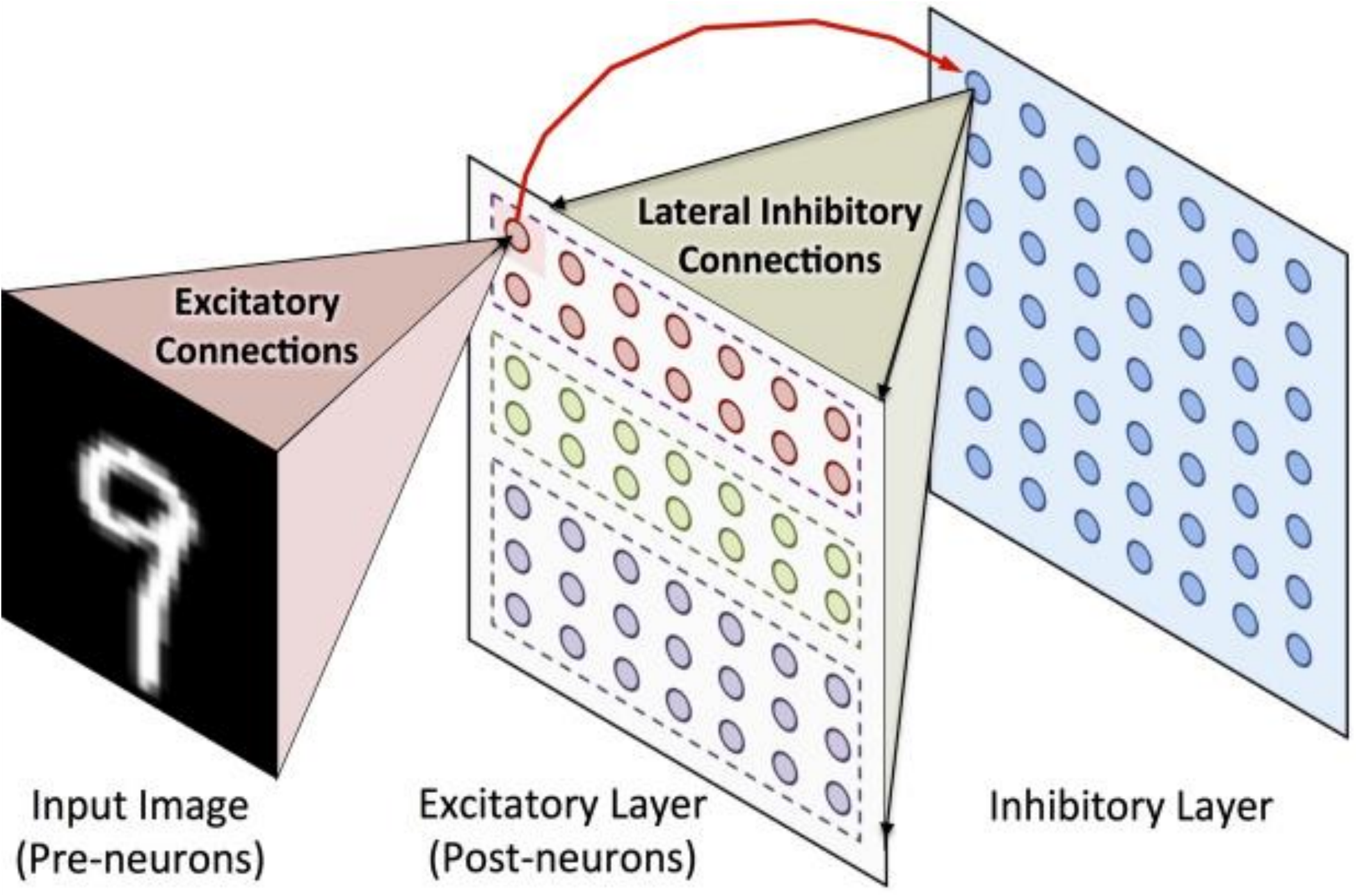}
\caption{Typical SNN topology used for handwritten-digit recognition application. It consists of an pre-neurons that correspond to the input image pixels. Pre-neurons are fully connected to the post-neurons in the excitatory layer. The inhibitory layer neurons inhibit the spiking activity of the excitatory neurons, except from which it receives the input, to ensure different patterns are learned across all neurons.}
\label{snntop}
\vspace{-3mm}
\end{figure}

To implement the dynamic learning algorithm under resource constrains using the proposed neuro-synaptic devices, an open source large scale SNN simulator, BRIAN \cite{brian}, was used. A typical two-layered SNN topology \cite{Diehl2015UnsupervisedPlasticity} was adopted shown in Fig. \ref{snntop}, for handwritten-digit recognition using the MNIST \cite{mnist} dataset. The input layer consists of a 2-D array of $28\times28$ neurons, each corresponding to a pixel in the input image. The rate of spiking activity of each of these neurons depends on the pixel intensity. The input neurons are fully connected to the excitatory neurons in the next layer. Further, each excitatory neuron connects to a corresponding inhibitory neuron which inhibits the spiking activity of all other excitatory neurons. This topology is known as \textit{Lateral Inhibition} and ensures that different input patterns are learned across different neurons.

The synapses connecting the input neurons to the excitatory neurons are trained to learn specific input patterns using the dynamic ASP learning rule \cite{aspanda,arxivasp}. This learning rule seamlessly integrates the traditional spike timing dependent plasticity (STDP) algorithm \cite{Clopath_2010} with synaptic weight decay. In STDP, the temporal correlation between the spiking activities of the pre- and post-neurons is utilized for training the synaptic weights. The synaptic weight is potentiated (depressed) when the pre-neuron spikes before (after) the post-neuron, according to the exponential dependence on the spike timing difference. However, ASP additionally allows for on-line incremental learning in non-stationary environments, without catastrophic forgetting, unlike STDP. In order to learn new class of patterns for a pre-trained network using STDP, the network will have to be re-trained using the entire dataset (including the patterns it has already learned) to avoid catastrophic forgetting. But ASP allows to learn new patterns on-line, without re-training the network with previous patterns. Some old patterns are replaced by new ones, and some are retained. The spiking activity of pre- and post-neurons were monitored to track the corresponding pre-/post-synaptic traces that were used to estimate the weight updates and decay rates of each synapse for the ASP learning rule. More details about the ASP algorithm can be found in \cite{arxivasp}.

\begin{figure}[t]
\centering
\includegraphics[width=3.2in]{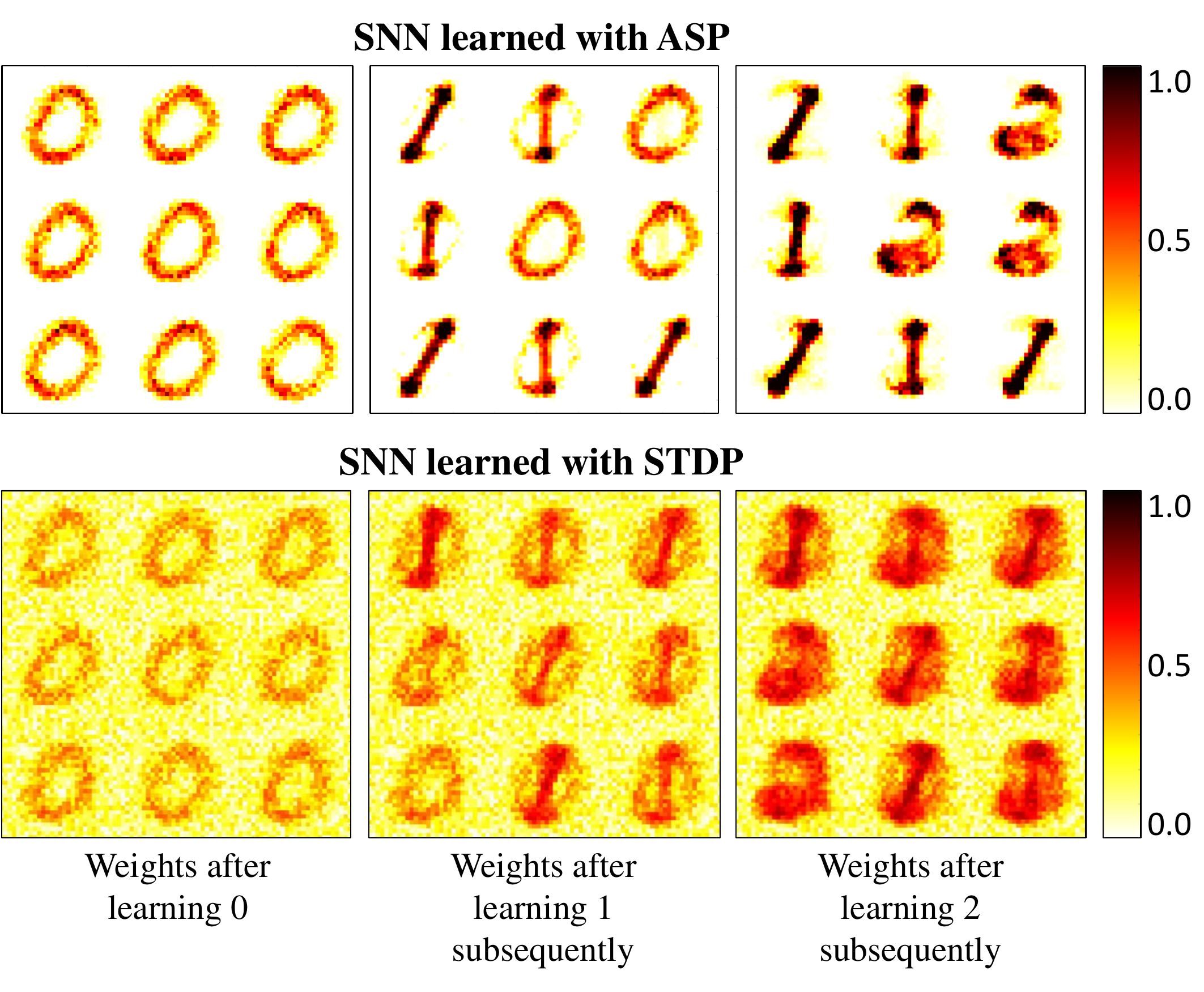}
\caption{The synaptic weights learned by the network (9 neurons) with ASP and STDP, in a dynamic environment where digits `0' through `2' were shown sequentially. The color intensity of the patterns represent the value of the synaptic weights.}
\label{fig:weights10n}
\vspace{-3mm}
\end{figure}

The neuron and synapse models developed for the proposed devices were incorporated into the BRIAN simulator using parameterized functional models and appropriate differential equations, extracted from the mixed-mode device simulation models detailed in Section \ref{sec:simframe}. First, to understand the advantages of ASP learning rule over traditional STDP, a small network with 9 excitatory neurons was sequentially presented with digits `0' through `2'. Fig. \ref{fig:weights10n} shows the synaptic weights learned by the network with ASP and STDP, respectively, with similar topology and learning environment. It can be observed that after 0's have been learned, ASP-learned SNN replaces a few 0's with 1's. Learning is stable because not all 0's are erased. Now, when digit `2' is presented, the older patterns (0's) are forgotten, while the latest patterns (1's) are retained. However, in STDP-learned network, new patterns overlap with previous patterns and the learning is unstable. This is because STDP has no mechanism to unlearn previously learned patterns. Interested readers are referred to \cite{aspanda, arxivasp} for various experiments that demonstrate the effectiveness of ASP over STDP.

\begin{figure*}[t]
\centering
\includegraphics[width=\textwidth]{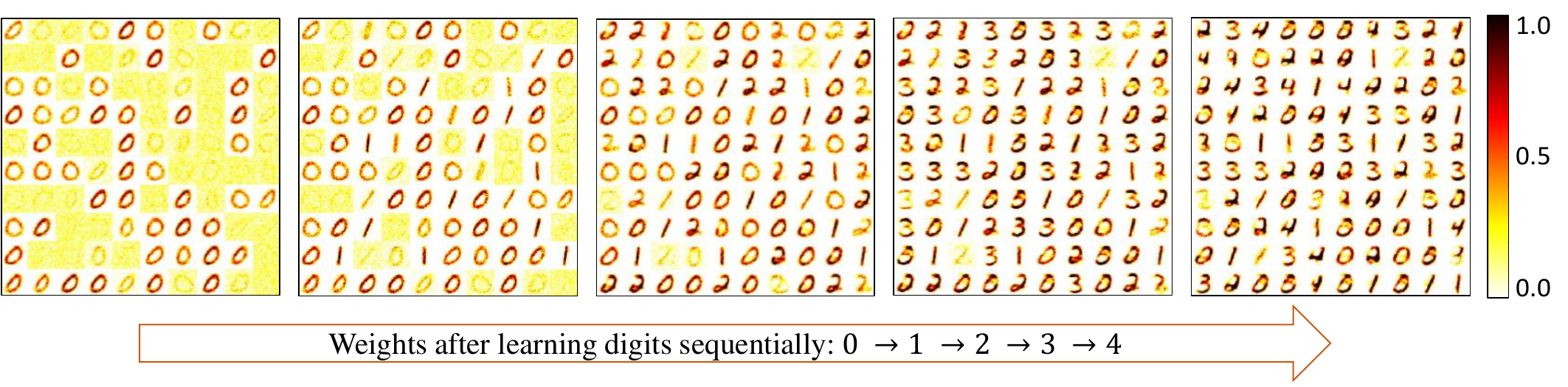}
\caption{The synaptic weights learned by the network after each batch of digits `0' through `4' were shown sequentially. The ASP learning rule enables the SNN to learn new patterns incrementally without catastrophic forgetting in a dynamic input environment. The color intensity of the patterns represent the value of the synaptic weights.}
\label{fig:weights}
\vspace{-3mm}
\end{figure*}

To test the feasibility of the proposed neuro-synaptic devices for implementing the ASP learning rule, we extend our toy network to an SNN with 100 excitatory neurons that was trained in a dynamically changing learning environment. Five classes of handwritten-digits, `0' through `4' from the MNIST dataset, were sequentially presented, with no digit re-shown to the network. First, all images of digit `0' were shown, followed by digit `1', and so on. Fig. \ref{fig:weights} shows the learned synaptic weights connecting the input neurons to 100 excitatory neurons, after each batch of digits were presented ($0 \rightarrow 1 \rightarrow 2 \rightarrow 3 \rightarrow 4$). It can be observed that the synaptic weights learn new patterns incrementally, and once all neurons are exhausted (after digit `2'), older patterns (digit 0) are un-learned and give way to newer patterns (digits `3' and `4'), without completely forgetting all the 0's. An average classification accuracy of $\sim$65\% was achieved for the SNN which learned the digits incrementally in a dynamic environment, similar to what was reported in \cite{arxivasp}. However, note that the classification accuracy increases with number of excitatory neurons ($\sim$95\% for 6400 neurons \cite{arxivasp}) due to the increased variety of input patterns that can be learned by the neurons.

\section{Conclusion}
Hardware implementations of neuromorphic systems are of paramount interest due to their efficiency in solving recognition and classification tasks. Towards that end, in this paper we propose a non-volatile leaky-integrate-fire neuron and a programmable synapse using the voltage driven bi-directional FM-DW motion. The FM-DW motion arises in response to the FE-DW motion of an underlying FE layer due to elastic coupling. The energy efficiency of the present proposal results from its intrinsic non-volatility and the pure voltage driven nature of the FM-DW movement. A mixed mode simulation framework consisting of micromagnetic simulation for magnetization dynamics and NEGF model for resistance of the MTJ was used to demonstrate the feasibility of the proposed neuro-synaptic device. Furthermore, the applicability of the proposed neuro-synaptic devices was shown for a typical handwritten-digit recognition application. The energy-efficient voltage-controlled behavior of the proposed devices make them suitable for emerging algorithms for dynamic on-line environments.

\section*{Acknowledgements}
The work was supported in part by, Center for Spintronic Materials, Interfaces, and Novel Architectures (C-SPIN), a MARCO and DARPA sponsored StarNet center, by the Semiconductor Research Corporation, the National Science Foundation, Intel Corporation and by the DoD Vannevar Bush Fellowship.

\bibliographystyle{IEEEtran}
\bibliography{ME_Neuron_TED_v3}

\end{document}